\documentstyle[12pt,epsfig]{article}
\newcommand{\be}{\begin{equation}}
\newcommand{\ee}{\end{equation}}
\newcommand{\ba}{\begin{eqnarray}}
\newcommand{\ea}{\end{eqnarray}}
\newcommand{\baa}{\begin{eqnarray*}}
\newcommand{\eaa}{\end{eqnarray*}}
\newcommand{\bb}{\begin{thebibliography}}
\newcommand{\eb}{\end{thebibliography}}

\newcommand{\lab}[1]{\label{#1}}

\begin{document}

\begin{center}
{ \large Elastic nucleon \\
        scattering  at small angles at LHC energies}

\vspace{.5cm}

S.V.Goloskokov, S.P. Kuleshov, O.V.Selyugin \\
BLTP,
Joint Institute for Nuclear Research, Dubna\\
\end{center}

\vspace{.5cm}

   The dynamical model of hadron-hadron scattering at high energies
  has been developed in [1-3] to describe particle interaction with the
  allowance for hadron structure at large distances; the model is based
  on the general principles of quantum field theory
  (analyticity, unitarity and so on) and takes into account basic
  information on the structure of a hadron as a compound system with
  the central region, in which the valence quarks are concentrated, and
  the long-distance region, in which the color-singlet quark-gluon field
  occurs.  As a result, the hadron amplitude can be  represented  as  a
  sum  of the central  and peripheral parts of the interaction [1,2]:
\ba
T(s,t) \propto T_{c}(s,t) + T_{p}(s,t),  \lab{tt}
\ea
 where $T_{c}(s,t)$   describes   the
interaction between the central parts of hadrons.
At high energies it is determined by the spinless pomeron exchange.
The   $T_{p}(s,t)$  is
the sum of triangle diagrams corresponding to
the interactions of the central part of one hadron  on  the  meson
cloud of the other. The meson-nucleon interaction leads to the spin flip
effects in the pomeron-hadron vertex [2].

The important feature of the model is the existence in the eikonal phase
of the term growing as $\sqrt{s}$
\begin{equation}
\chi(s,\rho)=\chi_0(s,\rho)+\sqrt{s}\; \chi_1(s,\rho),
\end{equation}
where the terms $\chi_i(s,\rho)$ logarithmically depend on the energy.
The term $\chi_1(s,\rho)$ looks like
\begin{equation}
\chi_1(s,\rho) \propto \int dz T_p^2 (s,r),
\end{equation}
where $T_p (s,r)$ is the peripheral part of the scattering amplitude in
the coordinate space.
So, the contribution growing as $\sqrt s$ to the eikonal phase is determined
by the peripheral meson-cloud effects. It has been shown that
this term becomes important for the energies $\sqrt s \ge 100 GeV$.

        The model provides  a self-consistent  picture  of
the differential cross sections and spin phenomena
of different hadron processes  at  high  energies [3,4].
Really, the parameters in the  amplitude  determined from a
reaction, for example, elastic $pp$-scattering, allow one to  obtain
a wide range of results for the elastic
meson-nucleon scattering and charge-exchange reaction
 $\pi^{-} p \rightarrow  \pi^{0} n$
 at high energies.
     The model predicts that at superhigh energies  the  polarization
effects of particles and antiparticles are the same [5,6].

  The model predictions of elastic proton-proton scattering at
$\sqrt{s} = 10$ and $20 \ TeV$ are shown on Fig.1 up to $|t| = 20 \
GeV^2$. It is clear that the differential cross sections at such
superhigh energies change their behavior and growth with increasing
energy.  This effect appears due to our peripherical term $T_p
 (s,t)$ which is determined by the pomeron interaction with the two
 pion cut and leads to a quick  growth of the total cross section as
 $\sigma_{tot} \sim (\log{s})^2)$ and growth of the differential cross
sections. It is more detailed shown on Fig.2.  It is evident that the
 growth of the size of proton increases the role of the
  periferical effects of superhigh energies.  At superhigh energies
 the eikonal at small impact parameters reaches its upper
 bound and the bound of unitarity is saturated (see Fig. 3). After that
 the differential cross sections change their behavior and begin
 growth at fixed transfer momenta.

Future FELIX experiments at LHC will give an excellent possibility to
 test the theoretical argument on the peripheral character of the
eikonal phase growth at superhigh energies determined by the
meson-cloud effects.

\vspace{.3cm}
\begin{center}
{\large                         Figure captions     } \\
\end{center}

Fig.1.  Predictions for pp - scattering
       -------------------- at  $\sqrt{s}=10 \ TeV$; \\
\phantom{.} \hspace{7.2cm}       - - - - - - - - - -   at
$\sqrt{s}=20 \  TeV$

Fig.2.  Predictions of $d \sigma/dt$ for pp - scattering at different
       $t$ \\
\phantom{.} \hspace{2cm} (full line - at $|t| = 2 \ GeV^2$;
	    long dashed line - at $|t| = 4 \ GeV^2$; \\
\phantom{.} \hspace{2cm} short dashed line - at $|t| = 8 \ GeV^2$;
	    dotted line - at $|t| = 16 \ GeV^2$;

Fig.3.  The form of eikonal $1-exp(-\chi(s,b))$
       -------------------- at  $\sqrt{s}=100 \ GeV$; \\
\phantom{.} \hspace{7.2cm}       - - - - - - - - - -   at
$\sqrt{s}=10 \  TeV$

\small{
\begin{center}
{\large            References     } \\
\end{center}
1. S.V.Goloskokov, S.P.Kuleshov, O.V.Selyugin, {\it Particles  and
              Nuclei} {\bf 18}, (1987) 39; \\
2.  S.V. Goloskokov, S.P. Kuleshov, O.V. Selyugin,
	  {\it Yad.Fiz.} {\bf 50}, (1989) 779.    \\
3.  S.V.Goloskokov, S.P.Kuleshov, O.V.Selyugin,
	   Z.Phys.C -Part. and Fields, {\bf 50}, (1991) 455;\\
4. S.V. Goloskokov, S.P. Kuleshov, O.V. Selyugin,
	  {\it Mod. Phys.Lett.} {\bf A 9}, (1994) 1207. \\
5.  S.V.Goloskokov, S.P.Kuleshov, O.V.Selyugin,
	  {\it Phys.of Atom. Nuclei} {\bf 58}, (1995) 1791.    \\
6.  N.Akchurin, S.V. Goloskokov, O.V. Selyugin,
	  JINR Comm. E2-92-78, (1997).
}
\newpage

%
  \vspace*{.5cm}
\epsfxsize=12.cm
\centerline{\epsfbox{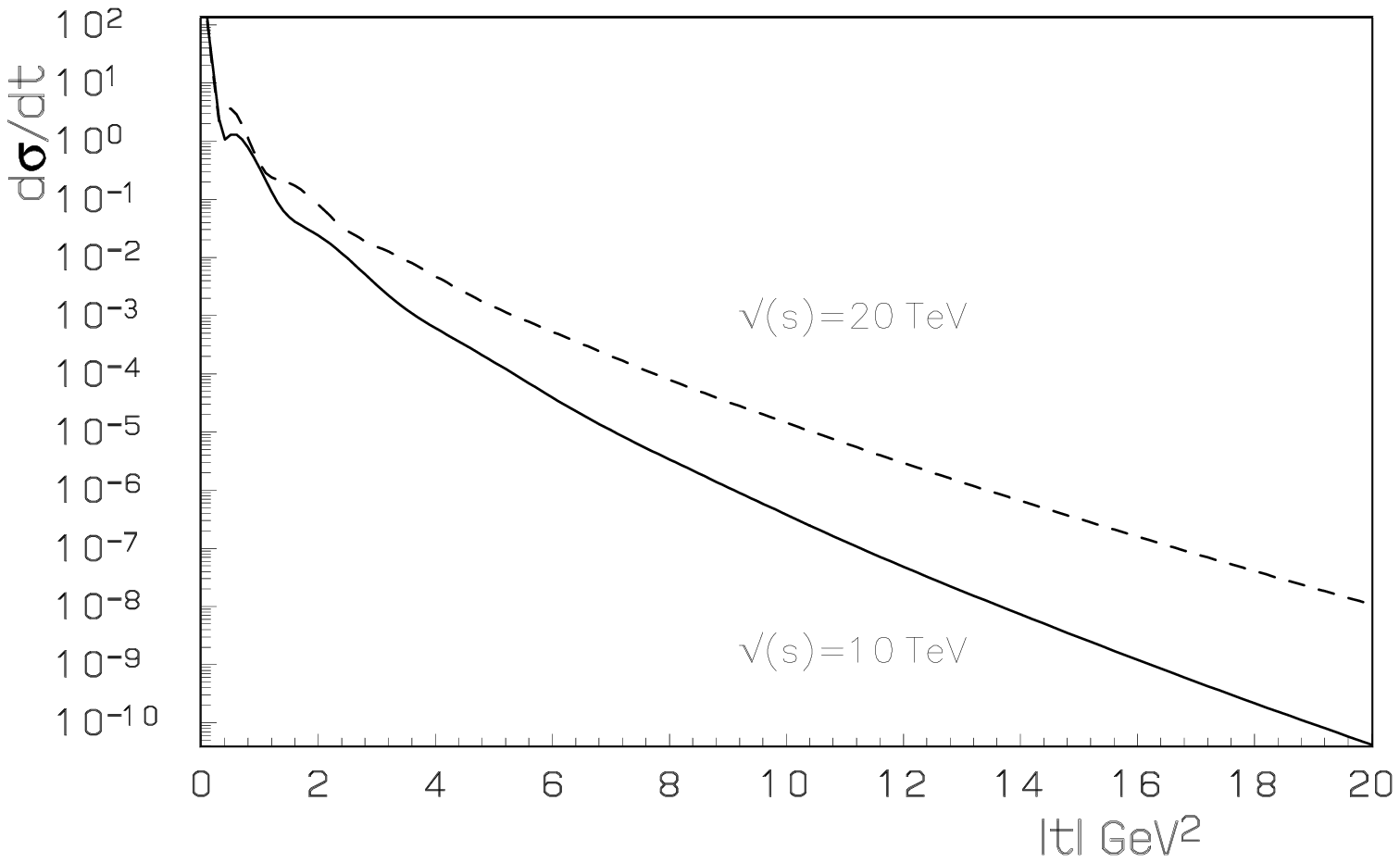}}
\begin{center}
Fig.1.
\end{center}
%
\epsfxsize=12.cm
\centerline{\epsfbox{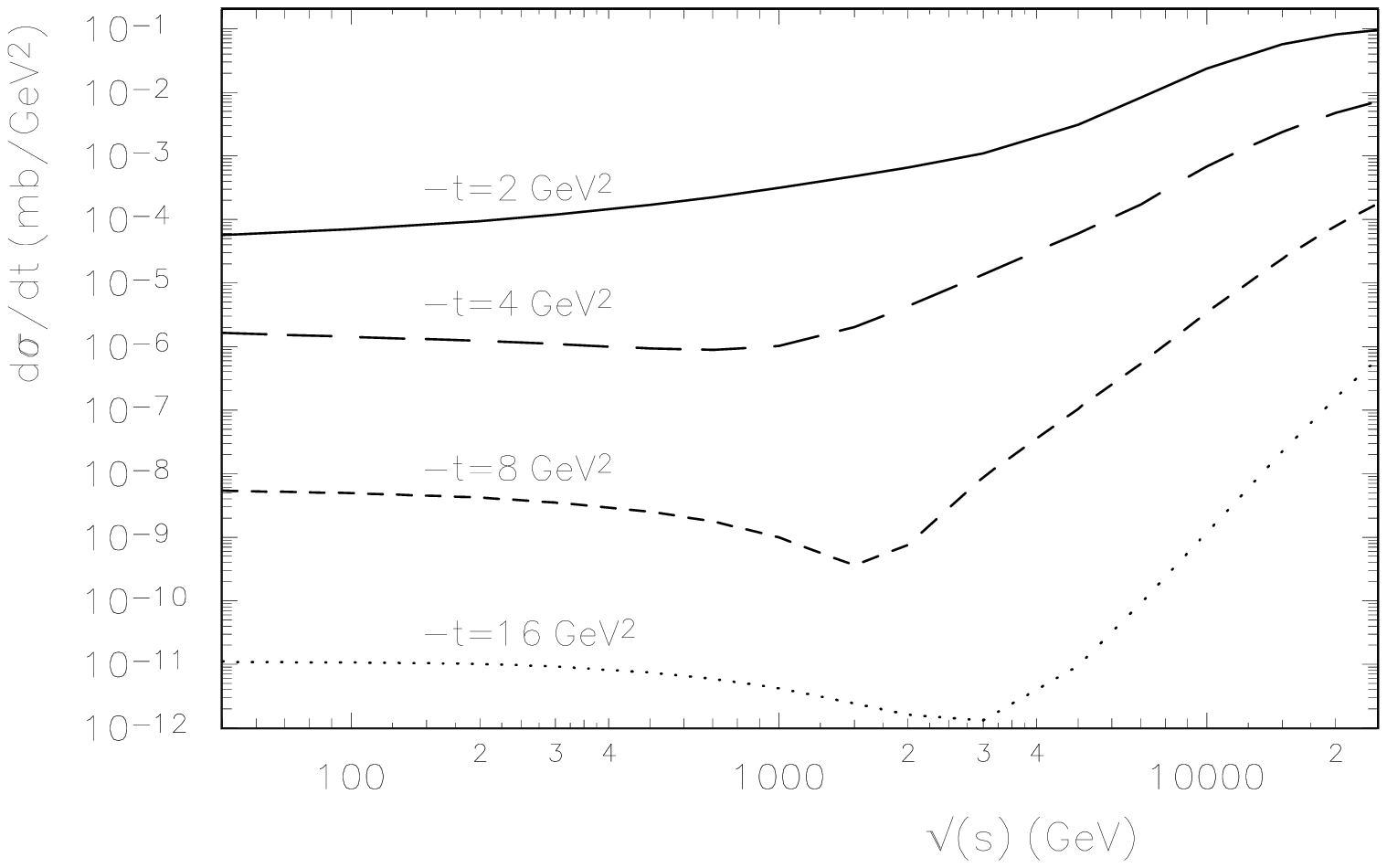}}
\vspace*{2cm}
\begin{center}
Fig.2.
\end{center}
\newpage
%
  \vspace*{.5cm}
\epsfxsize=12.cm
\centerline{\epsfbox{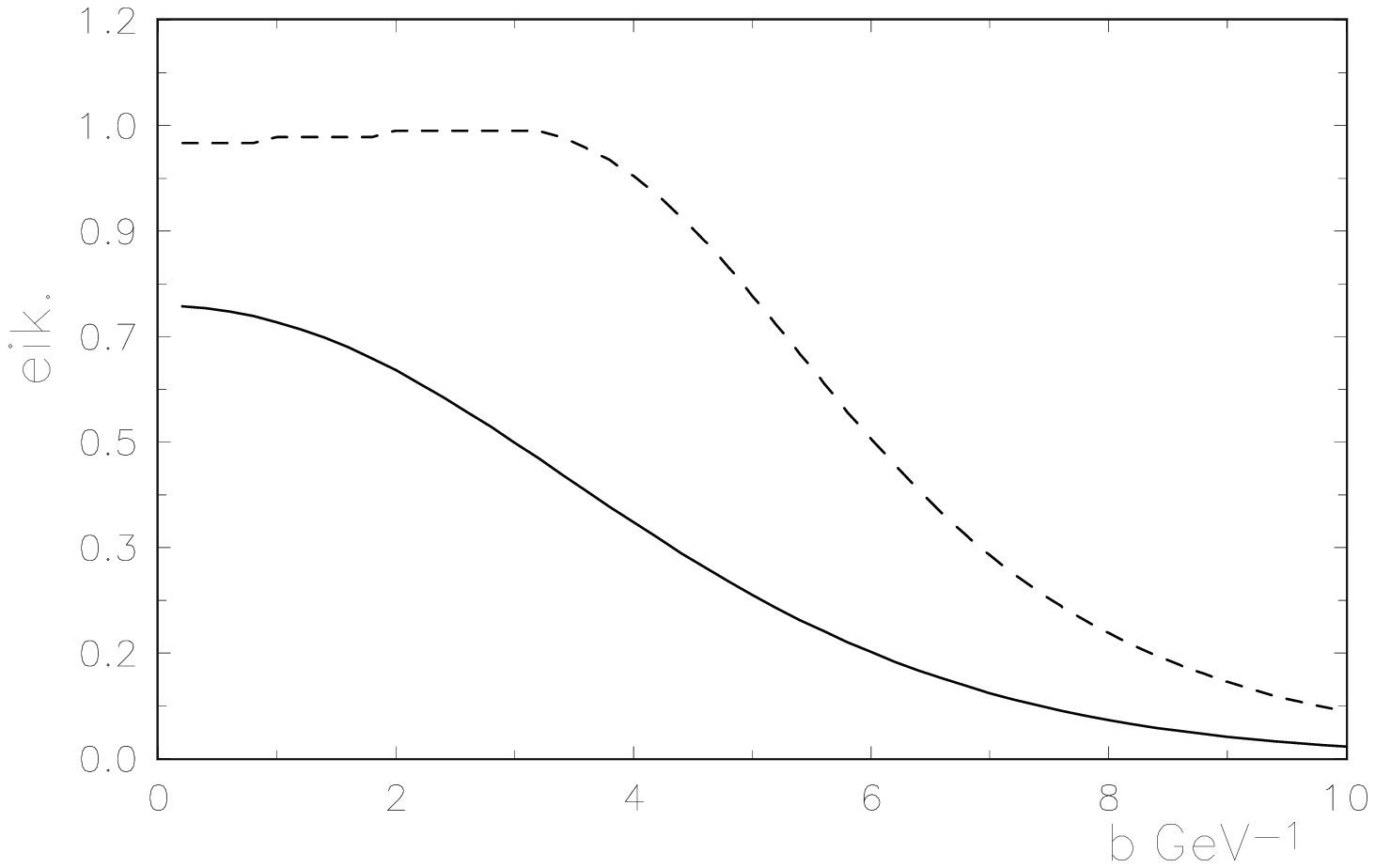}}
\begin{center}
Fig. 3.
\end{center}

\end{document}